# CREATING INTERACTION SCENARIOS WITH A NEW GRAPHICAL USER INTERFACE

First Céline Jost[1], Second Brigitte Le Pévédic[1], and Third Dominique Duhaut[1]

[1]Lab-STICC laboratory, Vannes, France

The field of human-centered computing has known a major progress these past few years. It is admitted that this field is multidisciplinary and that the human is the core of the system. It shows two matters of concern: multidisciplinary and human. The first one reveals that each discipline plays an important role in the global research and that the collaboration between everyone is needed. The second one explains that a growing number of researches aims at making the human commitment degree increase by giving him/her a decisive role in the human-machine interaction.

This paper focuses on these both concerns and presents MICE (Machines Interaction Control in their Environment) which is a system where the human is the one who makes the decisions to manage the interaction with the machines. In an ambient context, the human can decide of objects actions by creating interaction scenarios with a new visual programming language: scenL.

## I. INTRODUCTION

The human-centered computing is a multidisciplinary field because designing a software for human requires knowing human beings which is tried by philosophers, psychologists, biologists, therapists, neurologists, etc. There is currently no consensus. In the domain of computing sciences, there is an additional difficulty because schemes and formalization are needed to conceive application. But how can we formalize human beings if no formalization can design his/her functioning? Currently, there are a lot researches in various thematic as [17] emphasizes. It is possible to think that, in the future years, research will progress and little-by-little knowledge will increase. Work presented in this paper is based on a modular and adaptive architecture which can be enriched day after day called MICE: Machines Interaction Control in their Environment.

Moreover, there is the willing that the user participates to the interaction. To do that, some scientists work on visual languages. In [15], there is an analyze of some existing languages. Authors explain that a language is really important because nowadays computation time is lesser than communication time. Users spend a lot of time communicating with others, with the computer itself. They need languages to do that, to be understood by the computer. But all the presented languages in [15] and [8] are too complicated. Users are not expert in programming, so languages should be easier manipulated but it is the contrary, they offer complicated programming concepts. Indeed, these languages are graphical, but they stay complicated and users will never use it. On the contrary, Scratch [18][20], which is a language for kids, is really simple. But it is inappropriate here because it only allows to generate animations and does not offer parallel actions which is required in an ambient context. This language is really specialized but the willing of the presented work is to offer a more general language which can adapt all situations.

This paper presents a short state of the art of the human-centered. The following section introduces the context of the presented work, which is the Robadom project. Then MICEFrame the program which manages the interaction is introduced. The last section introduces experimentation and results about MICEFrame.

## II. THE HUMAN AT THE HEART OF THE SYSTEM

First, computers were made to calculate, they were human beings assistants. With the progress of technology, they became tools, useful for humans but never really autonomous. There is no trust in computers, we always refer to the human judgment. Robotic is subjected to the same story. At the beginning there were only industrial robots which were made to replace humans in automatic and complicated tasks as assembly, painting, packaging etc. Some years later, human beings were interested in the robot for others tasks which could directly help people. That was the birth of service robots which were developed to interact with people. It brought up researches on human-robot interaction where the robot should be a partner. These kinds of robots were not only evaluated for speed and accuracy but also for comfort. Some scientists focused on the fact that human factor is the most important and robots became companion robots where the purpose was to bring comfort to human and to act on mental health. These kinds of robot are only evaluated for comfort. In [21], Shibata explains the importance of a good human-robot interaction in a complete overview. It is important to take into account human-machine interaction more generally because there are non only robots, but also virtual characters. This research domain considers that a good interaction depends on the perception of users. That is why, they work on emotions, empathy... For example, scientists, like [1], tried to simulate a real interaction with the human by adding feedback cues in the behavior of the agent. Backchannel is really important because it gives empathy to the system and increase the chance to be accepted by the user. That explains why emotions is a big part of research, because without acceptability, the system will never be used. For example, in [2], a formalization has been found to allow an agent to express messages with speech and emotion, it increases the credibility of the agent. Emotion is so important that scientists extend existing models to add emotional support [7]. But it is important to know that each work is made for humans, there is a willing, according to [24], to add the human to the computational loop, the human has to be the core of the system, must even be a part of the program like think [6]. It involves two main problems : knowledge about humans and knowledge about the system. A lot of works tried to settle these problems. For example, in the so domain of nuclear, the TEPCO society tried to create a human-centered system in [12] to avoid accidents. Indeed, they noticed that the human responsibility was involved in most of incidents so they decide to work on a system which can avoid these kinds of problems. So, they studied all the past events to predict the future events. They studied the human behavior too. The success of the system is uncertain given the current events, but it allows to understand the importance to learn from the past. A good human-centered system should store past events, learn from it and should perfectly know the actors and the domain of the system. This idea is consolidated by [5] whose work consisted to design again an existing product taking account the human in the design. In [5], Chatterjee highlights that users need a personalized tool which gives a faithful feedback of their actions. Some researchers decided a lot of rules to create a human-centered system, like indicates in [3] and [16], but are they efficient and do people use them?

But, an efficient system is useless if it cannot understand users needs or if it cannot transmit messages to users. Thus, in the human-machine interaction, communication is fundamental because it allows exchange between the machine and the person. Cappelli et al [2] studied the human-robot interaction and more particularly all the possible modalities for communication. They conclude that the speech is the most efficient mean of communication because it is natural for human being. But they add it is more efficient if the system manages gaze and gesture too. The problem is that speech, gaze and gesture are hard to manage. Concerning speech, it is really difficult to have a good speech recognition [19] because the system needs to know all the words to be recognized. However, Kuhn et al [13] developed a system which recognizes any sentences. They create an interface which allows people to choose the TV program with several criterias like genre, name of actor, etc. They say that a remote control is not adapted to all person and that interfaces between humans and objects has to change, because technology change but everyone use old interfaces. This idea is in contradiction to [19] who says that new systems have to be integrated into existing system, transparently because person do not have to adapt to many change but to learn progressively.

To conclude, human-centered computing is an interdisciplinary active research area which has a lot of problematics. It is fundamental that the human plays the most important role in the interaction because she/he must be the controler. In [23], Talbert indicated that human-centered systems should complete the capabilities of humans and is adapted to human who should not learn how to use the system because systems have to be created from users needs Moreover, we live in an ambient world, so the systems should be dynamical to be faithful to the reality.

This state-of-the-art indicates that a human-centered system must generate a natural interaction (natural for human) with prediction: human has to be able to predict system behavior and system has to be able to predict human behavior. To do that, system has to keep past events in memory. A natural interaction also means emotion, empathy, good communication, personalization because each human being is different. Finally, a natural interaction also means that new systems have to be integrated into existing system to be transparent. That is why, this paper is focuses on ambient home context interaction.

III. ROBADOM PROJECT

Nowadays, one of the main concern is about the ageing of the population which causes numerous problems. Among them, there is the problem of their health. Indeed, it is estimated that in 2040, there will be three times more people over 85 years old than they are today. With this increase, the nursing homes will be definitely full and cannot accommodate others persons. Thus, the elderly will have to stay at home, but most of them need home helpers. The problem will move on them because it will not be possible for the home helpers to take care of more patients when their limits will be reached. And it is sure they will be reached! Consequences to the elderly will be a disaster, increasing their loneliness and depression.

The Robadom project [4], which is supported by the national research agency, focuses on this aspect. Its objective is to design a home care robot which will daily assist the elderly. This robot has several roles: (1) it will supervise and protect the patient, it is important to know if the person is well or not and it is important to react if the person has problems But the protection begins with prevention. The robot can analyze what happens around it and must indicate if there is a danger. Moreover, it is a help for doctors when it reminds to patients to take their medicine. (2) the robot is an assistant which manage the shopping lists, appointments etc. (3) It is an entertainment because, as a companion, the robot can speak with the person, can play with the person etc. (4) It is a social intermediary which can launch visual communication with the family, or give information about news etc.

The current patient are people with cognitive impairment, so the robot has to offer cognitive exercises. The project objective is to study the impact of such a robot on the elderly to know if it could be a solution to the ageing problems. The elderly is a concern in the entire world. For example, Heerink et al [11] tested the influence of a robot's social abilities on acceptance of elderly users but no correlation has be found between social abilities and technology acceptance. It is difficult to know what the elderly need exactly and what they want exactly. A study about game design for senior citizens [26] indicates that the elderly rejects computer because it can not replace a real person. It seems that these persons need to be useful, need to cultivate themselves and need to be connected to the society. The loneliness is the worst situation. That is why, the Robadom project wants to develop a system which fill the loneliness of the persons. Tefas et al [25] covered a part if this work by developing an application which can: (1) supervise the meal of the person and remind them to eat if they forget. (2) detect the facial expression of the person in order to analyze his/her emotion and express back an appropriate emotion.

Four partners are working on this project. The first partner is a firm which is in charge of the robot building. The Fig. 1. shows a preview of the future robot. It is composed of three parts: (1) a mobile base which can move in the house, (2) a computer which allows communication with the Internet, playing etc. (3) the trunk of the robot, which will be a penguin, designed in respect of T. Shibata study [21]. Indeed, the penguin is an unfamiliar animal and it increases the chance for the robot to be accepted by people.

The second partner is a hospital. There are therapists, psychologists and doctors who are studying the acceptability of the robot with their patient.

The third partner is working on sensors and is in charge of studying the sensor data to know the environment.

The last partner is specialized on emotional human-machine interaction and is in charge of building an architecture which merge others partners work.

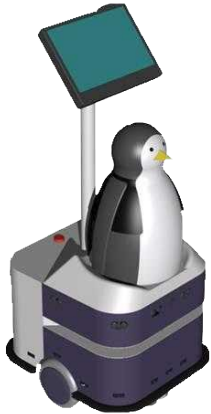

Fig. 1: Preview of the future robot of Robadm project

This project poses several problems. First, the robot has to understand speech and has to be understood by the human, who should feeling well and being listened to. Second, the robot has to know the environment and has to give a correct answer according to the situation. Finally, this answer has to use the modality of the human: speech, gesture, posture, etc. The robot has to be empathic and to show backchannel.

This paper focuses on the work of the fourth partner. The objective of this work is to provide an architecture which manages this paper the interaction between the human and the robot. More precisely, this paper presents MICE (Machine Interfaction Control in their Environment) which is an architecture which can manage the interaction between a human and a set of machines.

## IV. MICEFRAME INTERACTION MANAGING

MICE, which means Machines Interaction Control in their Environment, is a system which allows everybody to manage the interaction with a set of entities. An entity is *a physical or non physical object, which communicates with a human or a machine in the context of human-machine interaction. For example: a robot, a virtual character, either on a mobile or a television or a computer etc.*

### A. Objectives

MICE has two main objectives : (1) it allows the interaction between a set of input sensor and a set of output entities. Actually, the input sensors describe what happens in the environment and the output entities express the reaction of these events. (2) MICE offers an interaction programming tool which allows everyone to decide the action to do according to each event. Lee et al [14] almost did a similar work, their application has the same structure but there is a lack: the human cannot contribute to the interaction managing

### B. Overall organization

Fig. 2. shows the organization of MICE. The environment is composed of several sensors which can send raw events. A raw event is a data composed of:
- the name of the sensor which sends the event,
- the name of the event,
- the value of the event and
- a likehood coefficient, which is the certitude degree of the sensor concerning this event.

For example: the sensor "thermometer" can send ["thermometer", "temperature", "24", "100"]. That means that the thermometer indicates with 100% of certitude that the temperature is 24°C.

It is also possible to generation symbolic events from raw events. A symbolic event has the same format than a raw event and is computed by MICE. For example, a sensor of temperature sends a "temperature" events. The following rules show the creation if two symbolic events: **cold** and **hot**:

```
If (temperature < 15)
    Then (send_event(cold))
Else if (temperature > 27)
    Then (send_event(hot))
```

The advantage is that each user can define her/his own temperature limit. Symbolic events are personalized. Both types if events can participate to the creation of scenarios. For example:

```
If (cold)
    Then (put_heating_on)
Else if (hot)
    Then (put_heating_off)
```

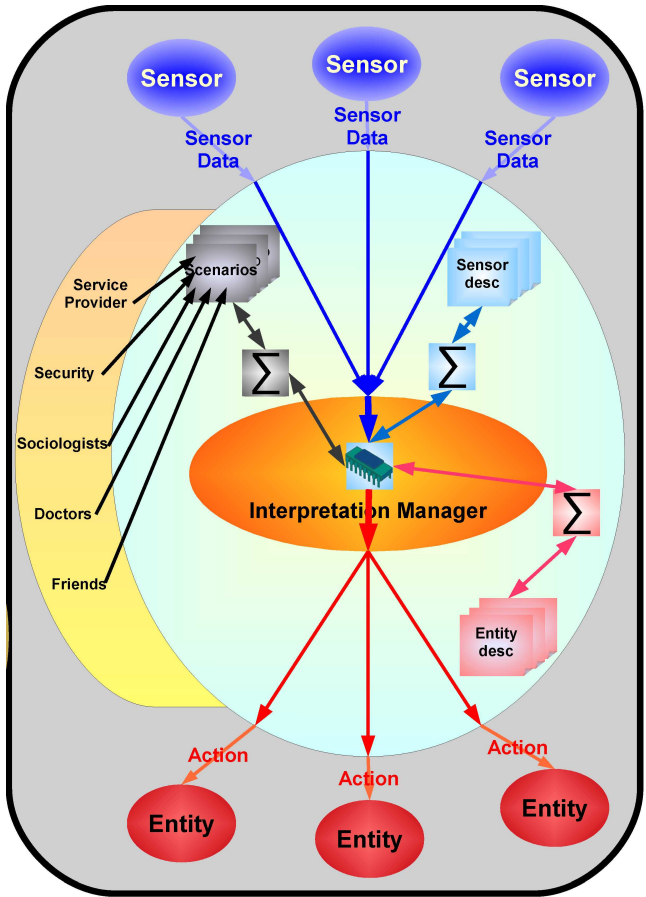

Fig. 2: MICE organization

Each event is sent to the Interpretation Manager which is the motor of the interaction computation. Final actions can be computed thanks to three types of files: entity descriptions files, sensor description files and scenarios. An entity description file is provided by the associated entity and contains the list of all available functions. A sensor description file is provided by the associated sensor and contains the list of all events that can be sent. A scenario file contains a scenario created by users, either friends, doctors, service provider or the patient her/himself. Finally, actions are computed and sent to the entities which have to do something.

### C. The scenario language: ScenL

The objective of ScenL is to be able to determine which actions to do, thus is is simple and does not look lije a classical programming language with class définition, inheritance, polymorphism... These concepts are not needed because it is possible to write scenarios with a simple language and because if the language is too complicated, users will never use it. Below, the ScenL grammar shows the simplicity of the language:

```
prog            ::instructions
instructions    ::instr ; { instr }*
instr           ::action |
                action_interrupt |
                repeat|
                while |
                parallel |
                conditional |
                event |
                timer |
                BREAK
action             ::ident.ident ( param$^{0/1}$ ) procedure_action
action_interrupt   ::° ident.ident ( param$^{0/1}$ ) °
integer_action     ::ident.ident ( param$^{0/1}$ ) integer_action
param              ::variable { , variable }*
variable           ::nb | integer_action | condition
repeat             ::nb *( instructions )
while              ::*[ cond ]( instructions )
conditional        ::[ cond ]( instructions ){ !( instructions ) }$^{0/1}$
event              ::< cond >( instructions )
cond               ::condition |
                   !( cond ) |
                   ( cond ) |
                   cond & cond |
                   cond | cond
condition          ::ident.ident ( param$^{0/1}$ ) boolean_action
parallel           ://( branches )
branches           ::instructions , { instructions }*
timer              ::WAIT( nb )

Terminal are written in bold and blue
ident is a string
nb is an integer
```

This language is structured by symbols contrary to others languages which are structured with key words like "if", "while"... Thus, programs are short and can be easier interpreted in real time. Despite its simplicity, ScenL offers some *conditional*, *loops*, *break*, *wait* like others languages which allows programing a big range of scenarios. A strengh of ScenL is *parallel* which allows to program several actions in the same time. Moreover, it contains *event* which allows to wait for a specific event to do something. This simple language is an alternative to factual, multi-agent and algorithmic programming.

To conclude, the language is simple, use basic concepts and is enough sufficient to program the interaction. But, to facilitate user learning, ScenL is manipulated through MICEFrame, its graphical user interface (see Fig. 3.)

### D. Programming some scenarios

MICE is associated to MICEFrame which is the programming graphical interface. Fig. 3. shows the organization of the frame. Area number one is a classical menu bar. Area number two represents ScenL conditions, that is the list of connected input sensors and their associated events. Area number three represents ScenL actions, that is the list of connected entities with their available functions. Area number four is the list of programming elements. This area represents the structure of the program and is static, while areas number two and three are dynamic. Users can create programs and save them as macro, which can be used in other programs. Area number five is the list of available macros.

To build a program, the user does drag and drop from elements at the right to the area number six. Fig. 4. shows a

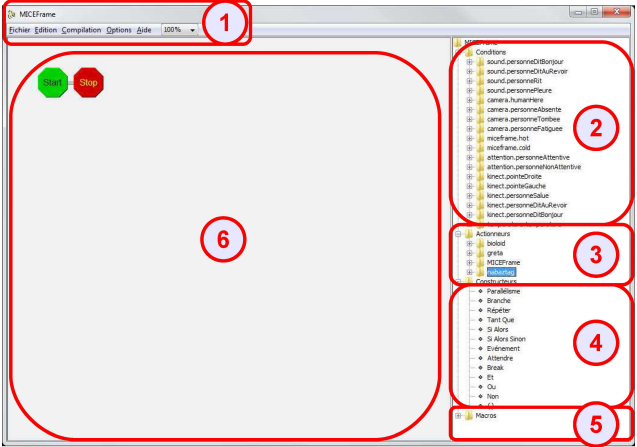

Fig. 3: MICEFrame structure

program example. It is written that the system waits for the

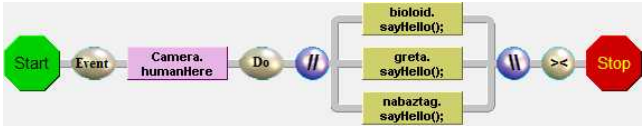

Fig. 4: Example of a ScenL program

*humanHere* event. When this event is detected, three sensors (bioloid, greta and nabaztag) execute their *sayHello* functions in parallel.

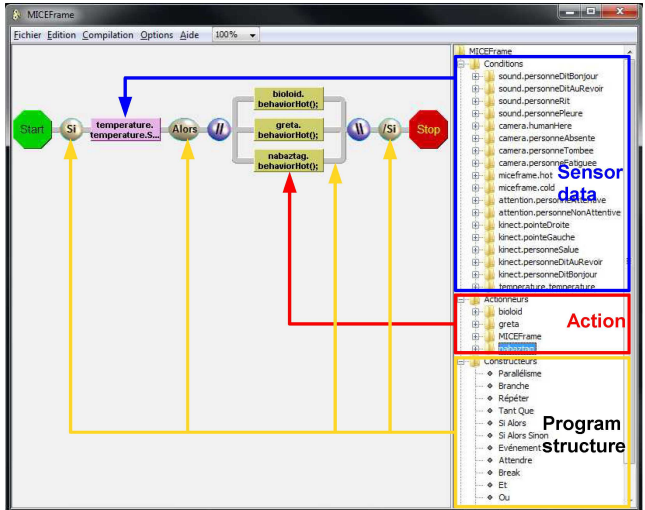

Fig. 5: Link between MICEFrame structure and ScenL example program

Fig. 5. shows the relation between MICEFrame structure and this example. Events are like conditions and actions are like the execution part of a program: *if **event** then **action***.

## V. EXPERIMENTATION FOR MICEFRAME EVALUATION

This preliminary experimentation allowed us to evaluate the quality of the MICEFrame visual programming language by adults that were not familiar with its use. We explored also if knowledge in computer sciences had an influence in the obtained results.

### A. Participants

Participants were 14 French adults without cognitive and physical disabilities (5 women and 9 men) between 23 and 58 years old (mean ± SD: 31.7 ± 5.3 years). Half of them (n=7) were considered themselves as expert in computer programming (group named *experts*), others had just little knowledge in this domain (group named *non experts*). All of them lived in Brittany, France. They gave us a verbal consent about their participation in this experimentation (data remained anonymous).

Notice that a man of 64 years old, with little knowledge in computing refused to do the task after receiving the experimental instructions and explanation of the graphical user interface. He explained that this kind of application was intended to younger generations, meaning he would never be able to succeed the task.

### B. Experimental design

The experimentation was made at participant's home or office (i.e. familiar place) and consisted of graphical user interface test. The participant was isolated with the experimenter in a room without any noise to make easier concentration.

*1) Equipment*

The MICEFrame graphical user interface (see Fig. 3) was displayed on a computer given by a computer engineer (i.e. named experimenter). The participant was seated face to the computer while the experimenter took place beside him/her. A chronometer was used to assess the latency to succeed or to fail the task.

*2) Experimental setting and data recording*

First, before the task, each participant was instructed by the experimenter as follows. Each participant was introduced to the advanced of technology and the difficulty for the human to control its computing environment. Thus, the experimenter proposed a visual programming language to resolve this problem. She explained then the different stages of the experimentation. At the end of this explanation, she asked if the participant agreed to take part to the experimentation.

Second, the experimenter began the experimentation by giving instructions. She introduced a part of MICEFrame, that were the programming area and the elements with correspond to functions, conditions and program structure. Since the experimentation focused on the concept, given examples were abstract. Indeed, if participants do not understand the meaning of the examples, they naturally try to understand the functioning of MICEFrame, that is not reciprocal: if you understand examples, you do not necessarily understand functioning of MICEFrame because you refer to the logical of examples.

After giving these instructions, the experimenter ensured that the participant had no question and started the task. Participant was asked to create her/himself a program from the following sentence: « If the person is felt then the bioloid calls the person while the nabaztag calls for help ». This sentence was also written on a paper next to the computer as an aide-memoire.

She/he was free to take her/his time to read this sentence and began when she/he was ready. The experimenter switched on the chronometer to start the task when participant first touched the computer mouse. Number of given helps and number of observed errors were registered. The experimenter stopped the chronometer when the participant either managed the task or asked to stop before the task was succeeded.

At the end of the experimental session, the participant answered a short questionnaire.

*C. Data collection and analysis*

*1) Questionnaire*

Seven questions were asked (Table I).

TABLE 1. ASKED QUESTIONS AND THE ASSOCIATED NUMBER

| Number | Question statements |
|---|---|
| 1 | Did you find the exercise was easy ? |
| 2 | Do you feel comfortable with the programming ? |
| 3 | Is it easy to understand the graphical user interface functioning ? |
| 4 | Is it easy to learn programming scenarios ? |
| 5 | Are you able to program other scenarios by yourself ? |
| 6 | If I give you MICEFrame and some objects for your home, will you « play » with MICEFrame or will you give up ? |
| 7 | Are you favorable (for or against) this kind of environment control ? |

Each question had its answer presented using a five-point Likert scale using five items: (1) strongly disagree, (2) disagree, (3) neither agree nor disagree, (4) agree and (5) strongly agree. Thus, for each question, the minimum score was 1 and the maximum score is 5. We had a global score that evaluated globally MICEFrame (i.e. total of these seven scores). Higher the score was, higher was MICEFrame positively perceived. We recorded all the answers with a sound recorder to have spontaneous comments and explanations in context.

*2) Simple observation data*

We gathered three additional data from each experimental session:

- Latency to succeed or to fail the task (in second): if the participant achieved the task, the experimenter stopped the chronometer when the participant seemed to think she/he was finished. If the participant gave up, the chronometer was stopped at the moment of the giving up.

- Number of given helps: either the participant asked questions to help her/him creating the program or the experimenter noticed that the participant was wrong (and would be blocked soon in the construction) and corrected her/him.

- Number of observed errors: the experimenter noticed that the participant was wrong. Generally, it was followed by a given help if the participant did not be able to correct.

*3) Statistical analysis*

Data analyses used Minitab 15© software. The accepted P level was 0.05. Data collected were score and time (latency in second). As our data were not normally distributed, we used a nonparametric statistical test called Mann-Whitney U tests and a Pearson correlation test [22] to study either knowledge differences in informatics or participant's age influenced the collected data.

## VI. RESULTS

All participants succeeded in the proposed task with a mean latency of 131.1±36.2 seconds (min-max: 51-284). The older the participant was, the higher was the latency to succeed the task (Pearson: rs=0.81, p<0.001; Fig. 6). Even if *experts* were quite faster in task resolution, difference with *non experts*' mean latency was not significant (X ±SD: 101.0±18.8 versus 161.3±44.3; Mann-Whitney U tests; n1=7 n2=7, U=42.0 p=0.201).

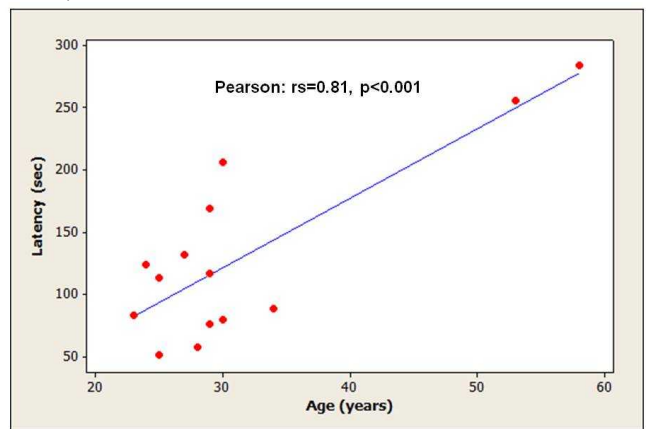

Fig. 6: Latency to succeed the task (seconds) in relation to participant's age (years). Pearson correlation, level of significance: p<0.05.

The mean total score of questionnaire was 32.1 ± 1.9 (maximal score was 35). No statistical difference was notices between *experts* and *non experts* (X ±SD: 33.6±0.9 versus 30.6±2.4; Mann-Whitney U tests; n1=7 n2=7, U=60 p=0.360). Mean scores of all questions were higher than 4, showing that MICEFrame was positively perceived. Here again, no statistical difference was noticed between *experts* and *non experts* (all Mann-Whitney U-tests, p>0.05; Fig. 7) except the question 4 (i.e. ease to learn programming scenarios) where *non experts* tended to be quite less agreed than *experts* (X ±SD: 5.0±0.0 versus 4.3.6±0.4; Mann-Whitney U tests; n1=7 n2=7, U=65 p=0.072; Fig. 7). The older the participant was, the lesser they felt comfortable with the programming (question 2; Pearson: rs=-0.626, p=0.017) and the lesser they found that learning programming scenarios was easy (question 4; Pearson: rs=-0.767, p=0.001). No other statistical difference was noticed according the participants' age (Pearson, p>0.05).

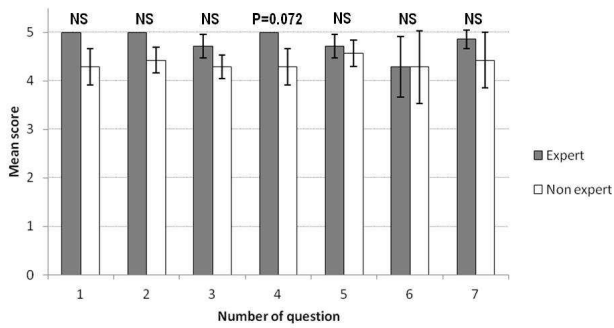

Fig. 7: Participant's answers to our seven questions about MICE according to their level of knowledge in computer programming (the question's statements were gathered in Table 1). Level of significance: p<0.05, NS: no significant; Mann Whitney U-test.

Six participants received some helps from the experimenter (X ±SD: 0.71 ±0.46) with a maximum of 2 helps. Nine participants made mistakes (X ±SD: 0.93 ±0.41) with a maximum of 2 errors. The older the participant was, the higher were the helps given (Pearson: rs=0.687, p=0.007). No statistical difference was noticed between *experts* and *non experts* (both Mann-Whitney U-tests, p>0.05; Fig. 8) and according to age for observed errors (Pearson: rs=0.183, p=0.532).

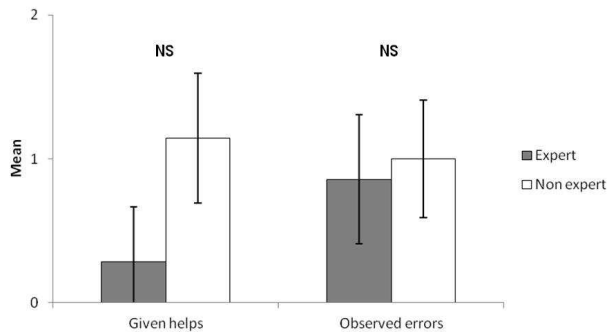

Fig. 8: Given helps and observed errors made by the participants according to their level of knowledge in computer programming. Level of significance: p<0.05, NS: no significant; Mann Whitney U-test.

Participants were free to give comments on the experimentation and on MICEFrame. We obtain two kinds of comments: technical comments and personal experiences comments. The main advantage of MICEFrame is that participants found it easy to use and funny. Their willing were to keep on using it, which is promising for a home integration of the system. The main disadvantage concerns the technical parts of the graphical user interface. Two participants indicated that the area with functions, conditions and programming structure was not enough intuitive and that is was maybe not accessible for everyone. One participant noticed that there was a lack of visibility of the graphical help. However, this point is debatable because two participants thought that MICEFrame was intuitive with a short adapation time. They enjoyed the fact that "once the functioning is understood, creating programs is really fast".

## VII. DISCUSSION

Our results showed that MICEFrame was positively perceived and that knowledge in computer sciences was not required to create scenarios.

### A. No significant difference between expert and non expert

Both *experts* and *non experts* achieved the task, none gave up, that was coherent with answer to question one. Moreover, they took, in mean, same time to succeed the task. All participants were able to create a scenario, after a short explanation, in less than five minutes. Thus, MICEFrame seemed easy to understand and to use and required a short learning time (here 5 minutes).

Concerning the questionnaire, *experts* and *non experts* answered differently at the question four. We hypothesized that *non experts* were not comfortable with this part of computing sciences and may not know their skills contrary to *experts* who are used to manipulate more complex systems. But, difference existed between mental representation (explored by questionnaire) and the reality (success in task realization) because *non experts* were able to create scenarios without showing great difficulties.

### B. Needs increase with age?

Our results showed that the older the participant was, the higher was the latency to succeed the task. We hypothesized this result may due to the fact that computing is young sciences (personal computers appeared forty years ago) and that older participant might not be as comfortable with computers as younger participants. Moreover, older participants received more helps than younger participants. Either it corroborated our hypothesis because older people were not able to be autonomous with computers compared to younger people who always knew it or it may be a bias of the experimenter (more prompt to help the older participant). It will be taken into account in future experimentation.

### C. An expert reticence?

During experimentation, *expert*'s attitude was totally different from *non expert*'s attitude. *Experts* showed a lack of commitment and tended to evaluate MICEFrame in place of *non experts*. Three of them were not favorable to MICEFrame because they preferred the idea of programming themselves, but they gave good notes because they were favorable for the non computer scientists.

### D. MICEFrame: a global enthusiasm

Our results showed that experimentation was positive, even if *non expert* had a tendency to limit their notation to 4, contrary to *experts* who easily gave 5. All *experts* had the same judgment although it was more contrasted from the *non expert* point of view. For example, answers to question six and seven were variable. Either it might mean that some people liked MICEFrame and other people disliked it or it might mean that *non expert* were not used to manipulate this kind of system and it was more difficult to imagine its application. This second hypothesis is confirmed by comments. Three women were really interested and wanted to keep exercises and found that programming could be funny with more experiences. Two participants imagined with excitement MICEFrame in their work environment as a solution to the difficulty of their tasks.

## VIII. Conclusion

This paper introduced MICE (Machines Interaction Control in their Environment) which allows a user to program the interaction between a set of sensors and a set of entities - *physical or non physical object, which communicates with a human or a machine in the context of human-machine interaction. For example: a robot, a virtual character, either on a mobile or a television or a computer etc.*

This paper introduced MICEFrame, which is the MICE graphical user interface. It allows to create some graphical programs, called scenarios, which represent the human-machine interaction. Human beings can control their digital environment and are the core of the system.

The preliminary evaluation of MICEFrame shows that it is positively perceived and that knowledge in computer sciences is not required to create scenarios. Our results indicate that MICEFrame is a pertinent solution which allows *non experts* to program and control their environment.

On the next stage, technical comments will be taken into account to improve the visibility of the graphical user interface. The right menu should be more intuitive. Moreover, a new experimentation will complete our current results. Indeed, the statistical analysis did not be able to make a correlation between age and knowledge. Thus, number of participants will be increase and the number of women and men will be the same in the two categories: *expert* and *non expert*. It will allow to study age and gender of participants.

Futures experimentation will include the elderly who are the target of the Robadom project. It means the experimentation will include people who do not have any knowledge in computing sciences.

## Acknowledgment

This work has been supported by French National Research Agency (ANR) through TecSan program (project Robadom n°ANR-09-TECS-012-02).